\documentclass[msmath, amssymb, twocolumn, aps, epl]{revtex4-1}

\newcommand{\Sc}{Schr\"{o}dinger }

\usepackage{amsfonts}   
\usepackage[dvips]{graphics}
\usepackage{graphicx}   
\usepackage{subfigure}
\usepackage{epsfig}
\usepackage{float}

\begin{document}

\title{Theoretical model of structure-dependent conductance crossover in disordered carbon}

\author{Mikhail V. Katkov}
\author{Somnath Bhattacharyya}
\email[Corresponding author:]{somnath.bhattacharyya@wits.ac.za}
\affiliation{Nano-Scale Transport Physics Laboratory and School of
Physics, University of the Witwatersrand, Private Bag 3, WITS 2050,
Johannesburg, South Africa }

\begin{abstract}
We analyze the effects of $sp^2/sp^3$ bond-aspect ratio on the
transport properties of amorphous carbon quasi-$1D$ structures where
structural disorder varies in a very non-linear manner with the
effective bandgap. Using a tight-binding approach the calculated
electron transmission showed a high probability over a wide region
around the Fermi-level for $sp^2$-rich carbon and also distinct
peaks close to the band edges for $sp^3$-rich carbon structures.
This model shows a sharp rise of the structure resistance with the
increase of $sp^3C\;\%$ followed by saturation in the wide bandgap
regime for carbon superlattice-like structures and suggests the
tuneable characteristic time of carbon-based devices.
\end{abstract}

\maketitle

\textit{Introduction}: Amorphous carbon ($a-C$) thin films have long
been considered as very important materials for understanding the
fundamentals of electronic structure \cite{beeman, robert, bred,
tersoff, galli,godet} and mechanical properties \cite{kugler,
dasgupta, mckenzie}. Depending on the preparation method, a very
diverse nature of microstructure can be formed ranging from
diamond-like to graphite-like carbon. Significant efforts have been
made to explain the optical properties of $a-C$ films based on the
estimated band gap and degree of disorder \cite{kelires, frauen,
tamor, stumm, ferrari, franch, davis, cherkashinin, carey}. There
are also numerous interesting results showing a structure-dependent
conductivity crossover in $a-C$ films \cite{alibart, shimakawa,
helmbold, sbprb}. However, a rigorous theoretical model of
electrical transport of $a-C$ films related to the carbon
microstructure is yet to be developed that can explain the
experimentally observed highly non-linear variation of the
conductivity with $sp^{2}$ bond concentration \cite{tamor, davis}.
In carbon the electronic properties can be strongly influenced by
bond disorder which depends on the $sp^{2}/ sp^{3}$ bond aspect
ratio and on the $sp^{2}$ cluster size. Although there is no well
defined structure for $a-C$ films they have been described in many
studies as a mixture of $sp^{2}$ and $sp^{3}$ hybridized bonds
\cite{kelires, frauen, tamor, stumm, ferrari, franch, davis}. A
number of spectroscopic techniques, for example Raman, optical
absorption and photoelectron spectroscopy, have confirmed this
assertion \cite{robert, dasgupta, franch, ferrari}. It was found
that the size of the $sp^{2}-C$ clusters greatly influenced the
band-gap \cite{robert}. The optical absorption edge was described by
a Gaussian plot based on the normal distribution of cluster size
\cite{mikulski}. Films with low $sp^2$ concentrations have also been
described as quasi-$1D$ polymer chains (e.g. trans-polyacetylene)
\cite{bred, franch, cherkashinin, carey, guin2, mikulski}. The
effect of $1D$ filamentary channels on electronic transport has been
observed in low-dimensional $a-C$ films and related devices
\cite{sbnmat}. However, no significant theoretical studies have yet
been undertaken to investigate the tunneling properties unlike in
$1D$ molecular structures \cite{onipko}. Based on previous studies
we believe that $a-C$ films can be described effectively as a
distribution of periodical alternation of the hopping energy and
constant energy term. Starting from previous experimental claims for
the variation of disorder with the energy gap \cite{tamor, ferrari,
franch}, we establish the trend of localization length in a wide
range of carbon films based on the calculated transmission
coefficients $T(E)$ and local density of states (LDOS). In this
Letter we propose a Gaussian disorder analogous to previous studies
\cite{tamor, ferrari, franch} but in a different manner which is
directly connected to the microstructure of $a-C$ films. Since the
nature of topological disorder has not been clearly understood from
previous works we briefly discuss this effect on electron
transmission at the end of this Letter.

\begin{figure}[tb]
    \centering
    \subfigure[]
    {
        \includegraphics[width=6cm]{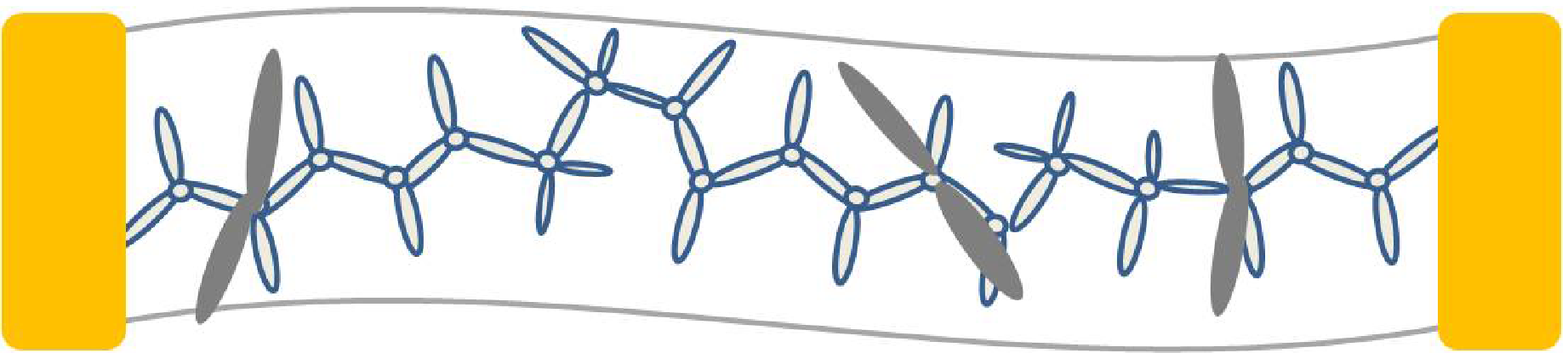}
        \label{carbon}
    }
    \\
    \subfigure[]
    {
        \includegraphics[width=3.8cm, height = 2.6cm]{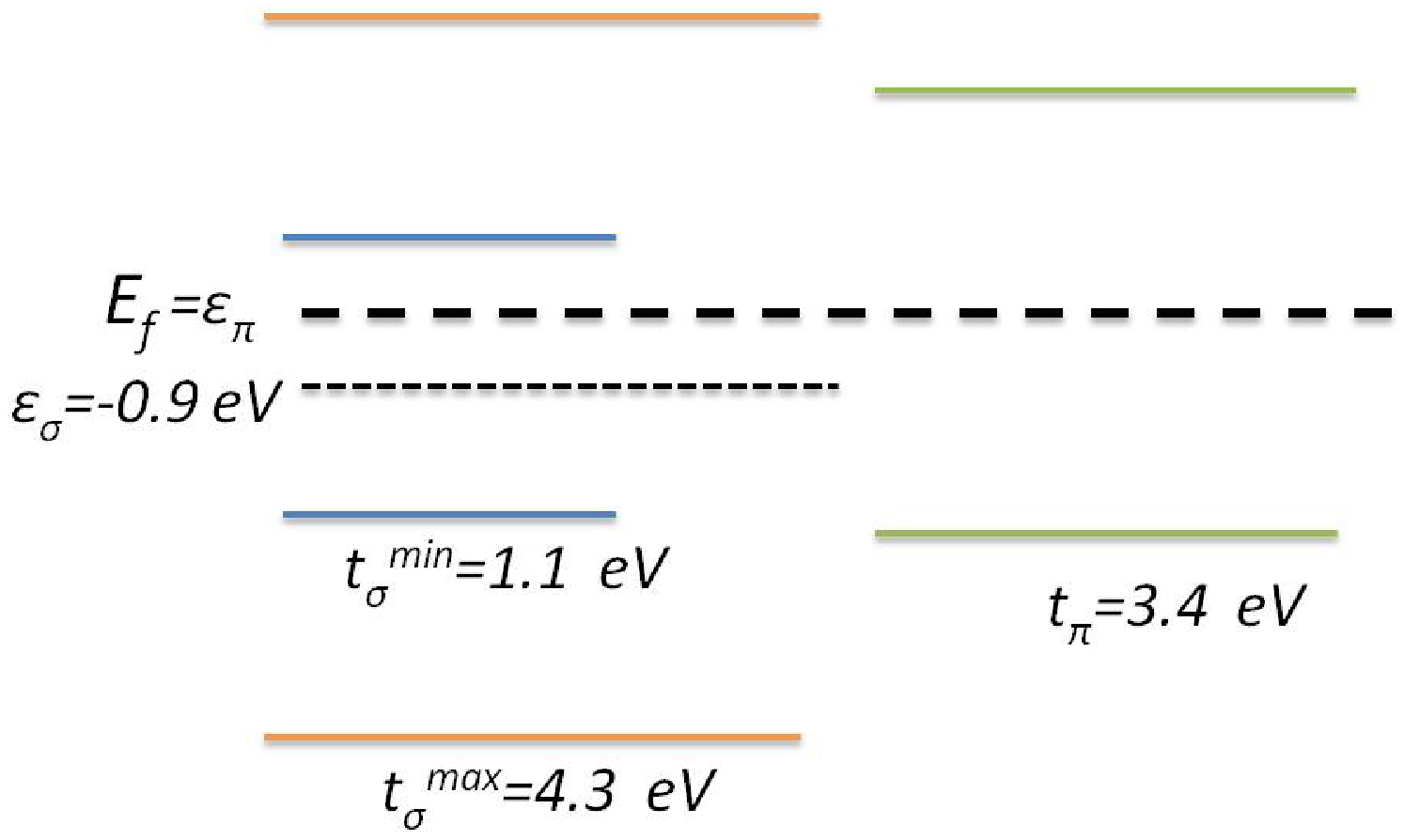}
        \label{diagram}
    }
    \subfigure[]
    {
        \includegraphics[width=3.6cm]{undist}
        \label{undist}
    }
    \caption{(a) Microstructure of $a-C$ consists of $\sigma$ and $\pi$ bonded carbon atoms. (b) The band diagram shows $sp^3$ hopping term alternation (red and blue lines online) and $sp^2$ hopping term level. The longer dashed line corresponds to $sp^2$ on-site energy, which is taken as the zero level, the shorter dashed line corresponds to $sp^3$ on-site energy. (c) $T(E)$ vs. incident electron energy ($E$) showed a number of quasi-bound states of mainly conjugated ($78\;\%$ $sp^2$) (dashed curve, red on-line) compared to pure saturated ($sp^3$) structure (solid curve).}
     \label{fig1}
\end{figure}

\textit{Proposed Structure}: We develop a quasi $1D$-dimensional
$a-C$ superstructure, which can be represented by a network of
narrow nano-ribbons with a length of ~$87\;{\AA}$ for the zigzag and
$75\;{\AA}$ for the armchair directions (Fig. \ref{carbon}).  This
structure is a quasi-one-dimensional representation of disordered
$a-C$, for similar examples see \cite{bred, Efstathiadis, Stephan}
which include some of the essential features specifically
$sp^{2}/sp^{3}$ clusters and bond angle distortions.There are 4
segments of the $sp^2$ structure with widths ranging from $1$ to
$10$ sites depending on the phase percentage. The superstructure is
a mix of saturated and conjugated ($sp^3-sp^2$) areas, corresponding
to $\sigma$ and $\pi$ bonds. The percentage of $sp^2$ carbon
structures determining the average $sp^2$ cluster size \cite{robert}
can be promoted, for example by nitrogen doping. The effects of the
lateral dimensions can be taken into account by considering
quantization of the wave vector in this direction and modifying the
hopping terms in the longitudinal direction.

\textit{Methodology}: We use a tight-binding Hamiltonian describing
electrons confined in $q-1D$ $a-C$ ribbons as
\begin{equation}
H=\sum\limits_n \epsilon_{n}c_{n}^{\dag}c_{n}-\sum\limits_n
t_{n+1,n}(c_{n+1}^{\dag}c_{n}+c_{n}^{\dag}c_{n+1}), \label{tham}
\end{equation}
where $\epsilon_{n}$ is an on-site energy, which can represent an
atomic energy as well as an external potential. $c_{n}^{\dag}$  and
$c_{n}$ are operators of creation and annihilation of electrons.
$t_{n+1,n}$ is a hopping term between site $n$ and $n+1$, which
takes into account nearest-neighbor hopping that can be different
from site to site reflecting the structural change. We seek the
solution of the \Sc equation \begin{eqnarray}\label{shred}
E\psi_{n}&=&\epsilon_{n}\psi_{n}-t_{n+1,n}\psi_{n+1}-t_{n-1,n}\psi_{n-1}
\end{eqnarray}
on the tight-binding lattice with the known eigenvalue $E=\epsilon_l
-2t\cos k_{l}\label{disp}$, where the unit cell length is fixed to
1. $t_l$ and $\epsilon_l$ are the hopping term and on-site energy,
respectively, corresponding to the left lead, which can be
interpreted as the Fermi energy ($E_f$), and $k_l$ is an incoming
wavevector. The boundary conditions imply that there is only an
outgoing plane wave at the right lead, while at the other sites in
the tight-binging lattice there are incident and reflected plane
waves. Starting from a unique outgoing wave on the right lead
$\psi_{n}=e^{ik_{r}n}$, we backpropagate this solution to obtain
$\psi_{1}$ and $\psi_{0}$ using Eq. (\ref{shred}). We get the
incoming wave amplitude $A$ and define
$T(E)=\frac{1}{|A|^{2}}\frac{\sin k_{l}}{\sin k_{r}}$, (outgoing
wavevector, $k_r$). We calculate the transmission coefficient as a
function of incident electron energy, where the transmission
coefficient is the ratio of the outgoing to incoming probability
current (since $k$ vectors may be different for incoming and
outgoing waves). Thus, the problem is different from that of finding
the eigenvalues as the transmission coefficient is evaluated as a
function of energy. The resonance energies determined from
transmission maxima correspond to eigen-energies of the system.

\textit{The Carbon model}: In this model a Hamiltonian with
alternating values of the resonance integral produces conduction and
valence bands as well as a band gap \cite{onipko}. For the saturated
region we introduce the resonance integral alternation by the
hopping terms $t^\sigma_{min}$ and $t^\sigma_{max}$. In the presence
of the conjugated compound the effective gap decreases because of
the electron resonance at energies lying inside the $sp^3$ phase
band gap. The hopping term $t_\pi$ corresponding to the $sp^2$ phase
experiences distortion due to the structural disorder, which
reflects the difference in the bond length via the deformation
potential. Topological disorder corresponding to the difference in
the cluster size affects the resonance conditions via the boundary
conditions for quasi-bound states \cite{franch}. The disorder of the
$sp^3$ phase is neglected due to its minimal effect on the resonant
electron transmission within the gap \cite{robert}. The broadening
effects are mostly due to disorder in the well. As a matter of fact
the resonant states have a larger amplitude inside the wells in this
range of energies. The leads on the right and left sides are taken
as ordered narrow $1D$ graphite-type strips. The incoming electron
energies are given with respect to the $\pi$ electron ($sp^2-C$)
on-site energy $\epsilon_\pi=0\;eV$. Therefore, the position of the
valence and conduction $sp^3$ phase is not symmetric due to the
different on-site energies for $\pi$ and $\sigma$ electrons and has
a shift corresponding to the difference in these energies, as shown
in Fig. \ref{diagram}. The hopping between $sp^3$ and $sp^2$
hybridized $C$-atoms is extremely significant due to the non-planar
geometry of the $sp^3$ structure \cite{kugler}. We use the following
tight-binding parameters: $sp^3$ on-site energy
$\epsilon_\sigma=-0.9\;eV$, $sp^2-sp^2$ transfer integral:
$t_{2-2}=3.40\;eV$, $sp^2-sp^3$ transfer integral
$t_{\sigma-\pi}=1.37 \;eV$, $sp^3-sp^3$ transfer integrals
$t_\sigma^{min}=1.10\;eV$ and $t_\sigma^{max}=4.30\;eV$
\cite{onipko,hjort}. Fig. \ref{diagram} shows the levels of hopping
terms along with on-site energies for atoms in $sp^3$ ($\sigma$) and
$sp^2$ ($\pi$) structures.

\begin{figure}[b]
\centering

 \centering
    \subfigure
    {
        \includegraphics[width=5.5cm]{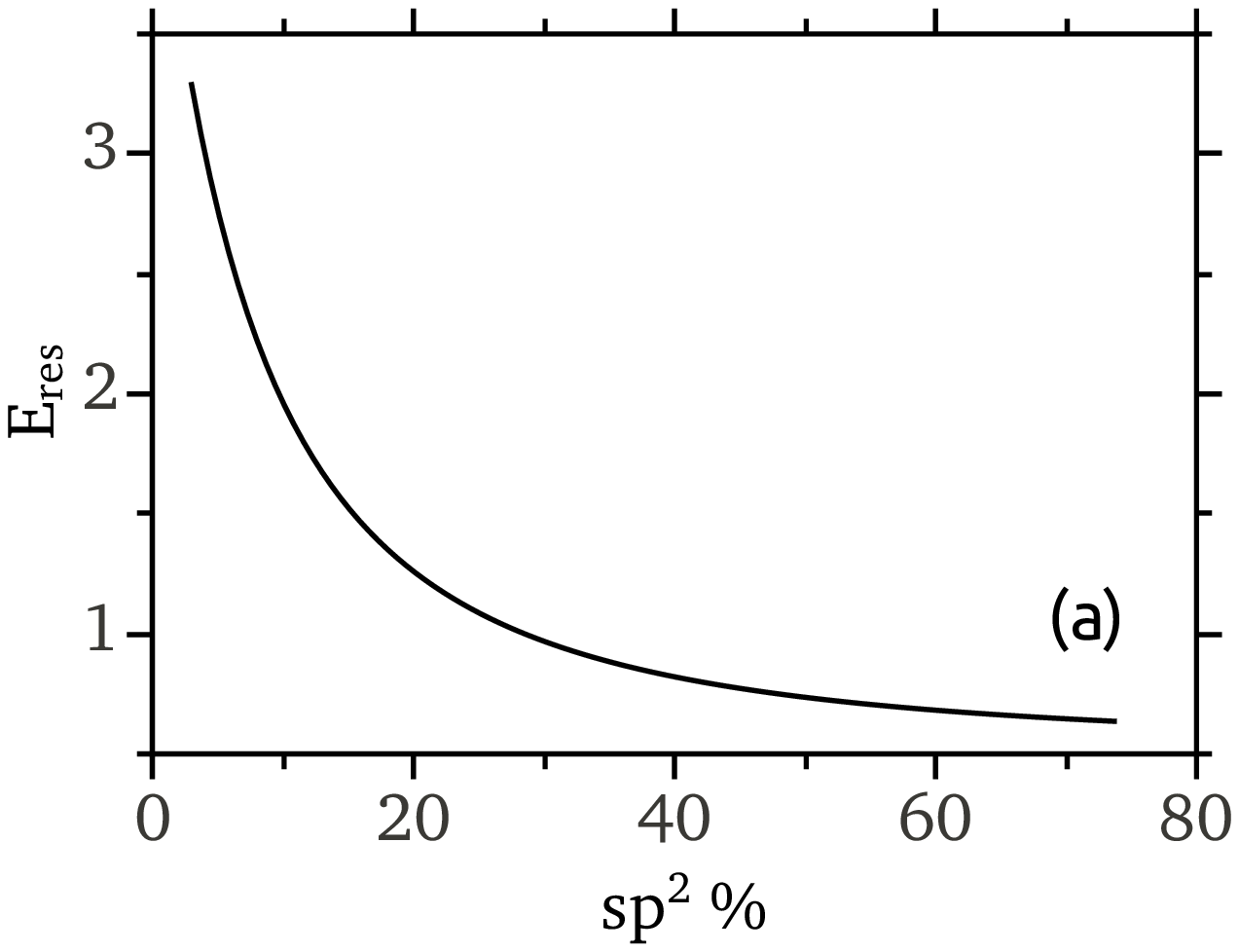}
        \label{res_per}
    }
    \subfigure
    {
        \includegraphics[width=5.5cm,  height = 4.5cm]{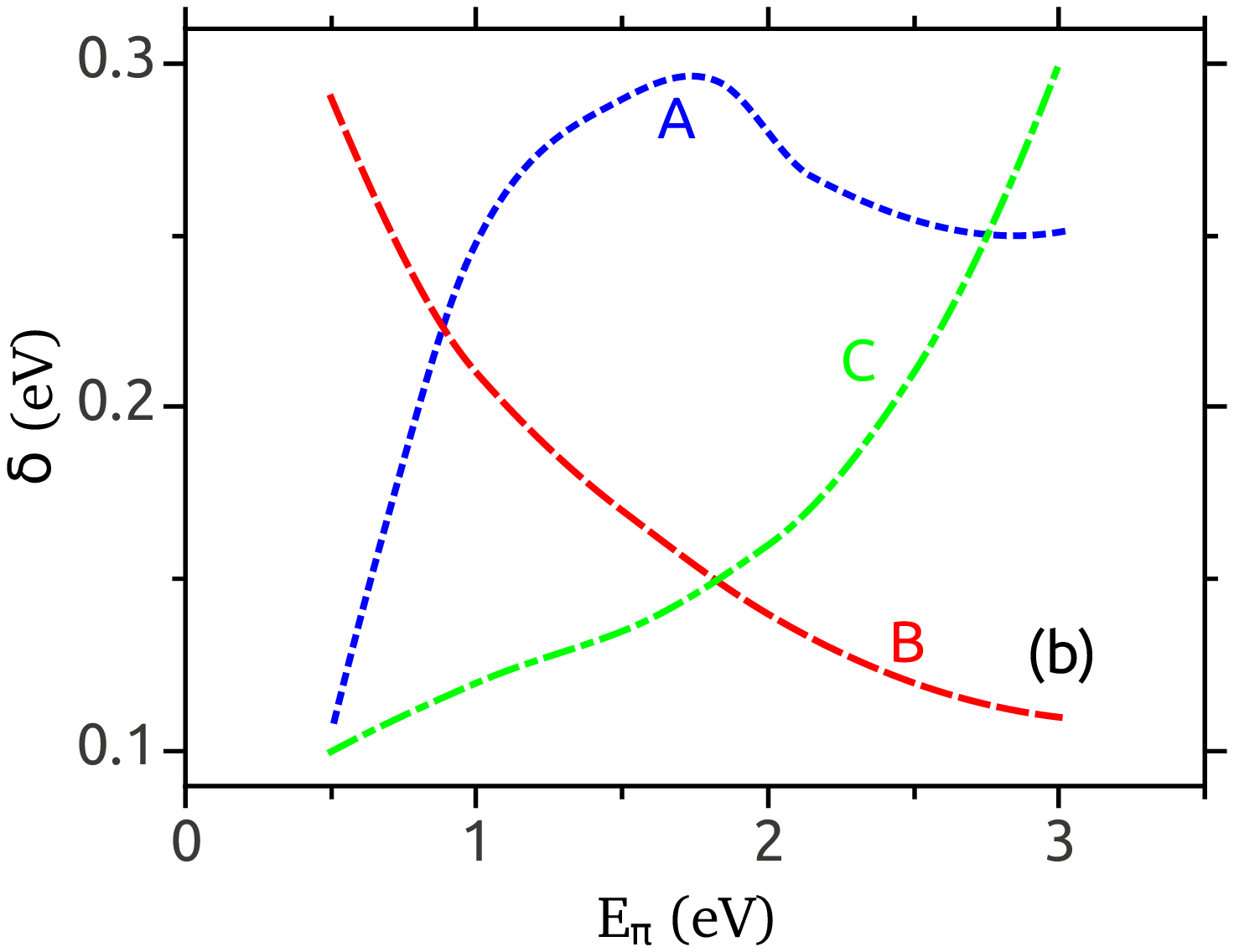}
        \label{delta}
    }
\caption{(a) First resonance peak from the zero level $E_{res}$ as a
function of $sp^2$ phase percentage. (b) Hopping disorder parameter
$\delta$ of the $sp^2$ region for cases A, B and C as a function of
$E_{\pi}$, where $2E_{\pi}$ is the peak-to-peak energy spacing of
bonding and antibonding states.} \label{pi_delta}
\end{figure}

Fig. \ref{undist} shows $T(E)$ of a pure saturated structure and an
ordered structure with $78\;\%$ of $sp^2$ bonding concentration
where the zero level corresponds to $sp^2$ on-site energy. The
numerically calculated band gap and band widths of the saturated
structure are $E_\sigma^g=6.4\;eV$ and $\Delta
E_\sigma^{c(v)}=2.2\;eV$, respectively. For the mixed structure,
there are several narrow peaks within the $sp^3$ band gap
corresponding to quasi-bound states of the $sp^2$ ``wells". The
peaks are very sharp with a high peak-to-valley ratio. In the case
of a symmetric structure, $T(E)$ always reaches 1 \cite{onipko}.
When the concentration of $sp^2$ structure is high, the peaks are
close to the $E_f$. On the other side of the concentration limit
(low $sp^2$ phase), the peaks are close to the band edges of $sp^3$
structure. Fig. \ref{pi_delta}(a) shows the first resonance peak
position $E_{res}$ (from the zero energy level) as a function of
$sp^2$ phase percentage, which has an inverse square dependence on
the percentage. Such a dependence reflects a linear increase of the
average $sp^2$ cluster size and a quadratic decrease of the energy
associated with the resonance.

\textit{Effect of Disorder}: There is a significant effect of
structural $sp^2$ disorder on the transport properties of the
$sp^3-sp^2$ carbon system. Disorder is an inherent micro-structural
property of $a-C$ films. It has been suggested that the density of
the bonding and anti-bonding ${\pi}$ states could be represented
with a Gaussian distribution \cite{franch, dasgupta, mikulski}. In
particular the Urbach energy was described as a function of the
width of the ${\pi}$ DOS peak at $E_{f}$ although the link between
disorder and the broadening parameter was not clear. The disorder
parameter ${\delta}$, which takes account of bond length (angle)
distortions, was calculated previously \cite{franch}.
Micro-structural disorder is broadly separated into structural and
topological disorder although the later only applies for a large
$sp^2$ clusters.
 In this Letter a typical hopping disorder parameter $\delta$ for a Gaussian distribution of
the hopping term $t_\pi$ is shown in Fig. \ref{pi_delta}(b) for 3
cases, A, B and C as a function of $E_{\pi}$, which covers the major
possible behaviors of disorder in carbon structures. Models B and C
are suggestions for the behavior of the disorder parameter based on
the assumption that the disorder parameter can either increase or
decrease with the $E_{\pi}$ corresponding to either $sp^{3}$ rich or
$sp^{2}$ rich structures. The values of the disorder parameter for
model A $\delta$ can be extracted from a number of experiments
(e.g., Raman $G$-peak linewidth) \cite{tamor, franch, ferrari} and
we believe it is the result of two competing processes (B and C). On
this basis, we have determined the behavior of the disorder
parameter as a function of $E_{\pi}$. We therefore focus on model A
as it corresponds to the physical nature of $a-C$ systems. A number
of $sp^{2}-C$ percentages were chosen to cover a wide range. The
corresponding $E_{\pi}$ values (and hence corresponding values for
${\delta}$) were determined based on the relationship between
$sp^{2}-C\;\%$ and $E_{\pi}$. We found that the values of $E_{\pi}$
were very close to the values of $E_{res}$. Here $2E_{\pi}$ is the
peak-to-peak energy spacing of bonding and antibonding DOS above and
below the $E_f$ \cite{franch}. For case A, the disorder reaches its
maximum at about $E_{\pi}=1.9\;eV$, followed by a slight decrease.
This is an attribute of the structural relaxation, which occurs
(also for case B) when $sp^2$ clusters become relatively large
\cite{robert,bred}. For case C the disorder increases sharply with
$E_{\pi}$ corresponding to the rise of $sp^3$ phase content. In our
further calculations we associate $E_{\pi}$ with the energy of the
first (from zero) resonance peak, $E_{res}$. The proposed non-linear
graph (case A) showed a significantly different trend of $T(E)$ and
localization length compared with the two other cases where a
continuous decrease (B) or increase (C) of $\delta$ with $E_{\pi}$
is considered.

\begin{figure}
\centering
\epsfig{file=01.eps,width=0.75\linewidth,clip=}
\epsfig{file=02.eps,width=0.75\linewidth,clip=}
\epsfig{file=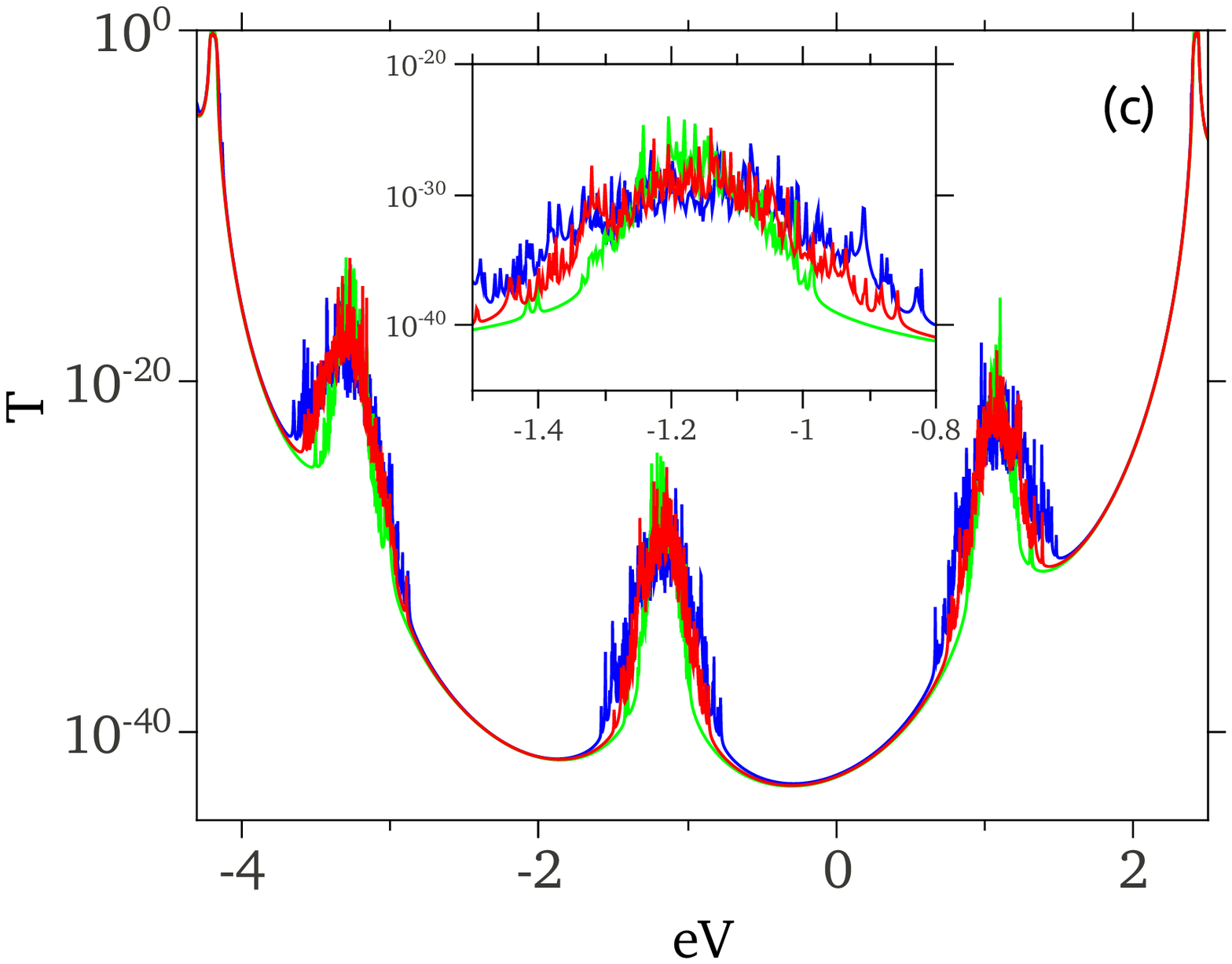,width=0.75\linewidth,clip=}
\epsfig{file=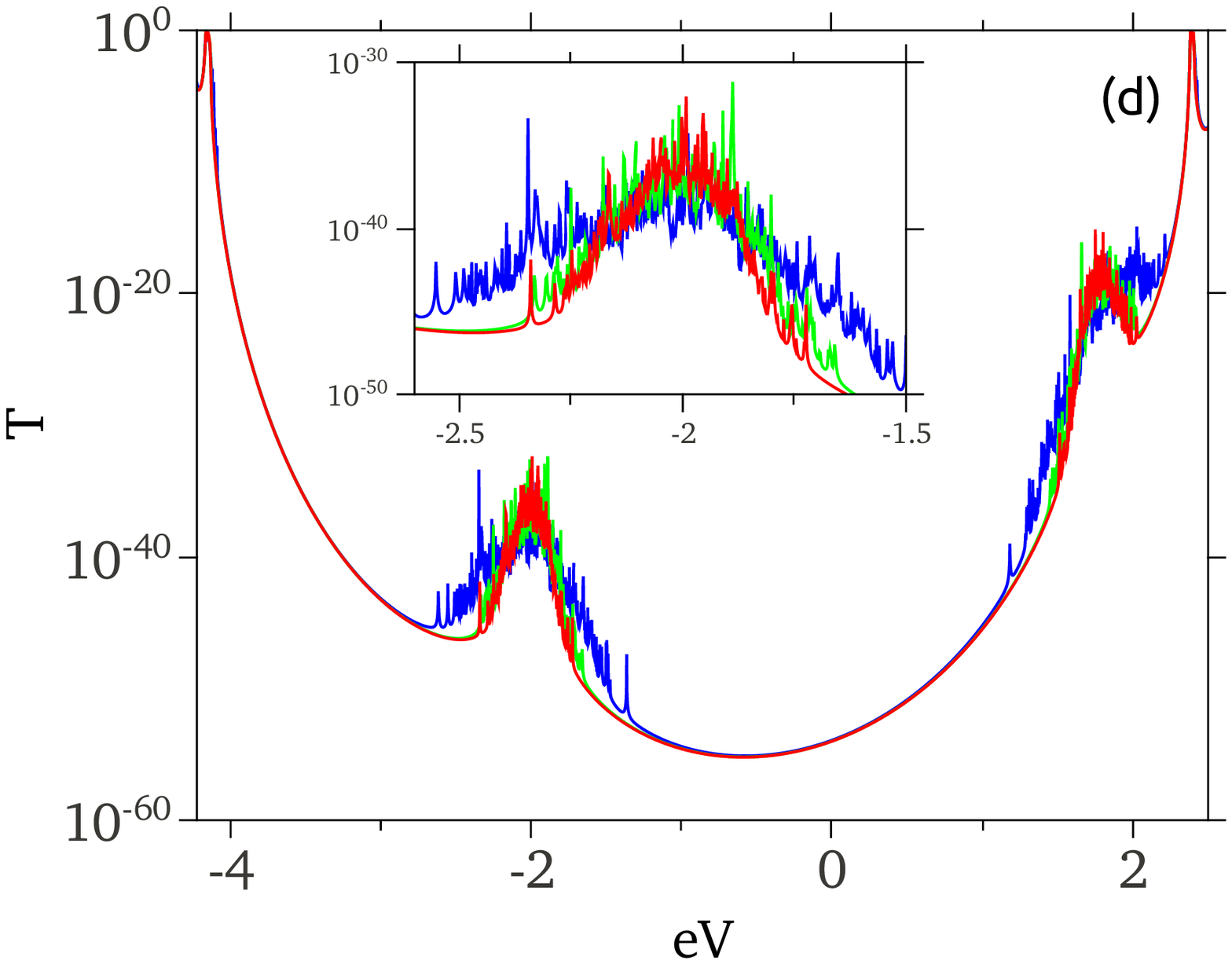,width=0.75\linewidth,clip=}
\caption{$T(E)$ vs. $E$ for the following $sp^2$ bond concentration:
(a) $72\;\%$ ,(b) $42\;\%$ ,(c) $24\;\%$, and (d) $7.8\;\%$. In
subfigures (c) and (d) 3 curves (in color online) correspond to 3
cases of hopping disorder parameter $\delta$ dependance on $E_{\pi}$
as shown in Fig. \ref{pi_delta}(b). The tight-binding parameters are
the same as that in Fig. \ref{undist}. The peaks for Model A showed
a broader feature than that of case B and C, shown in the insets of
Fig. (c) and (d).} \label{trans}
\end{figure}

In Fig. (\ref{trans}) (a) we show $T(E)$ for $73\;\%$ $sp^2$ bond
concentration, which corresponds to $2E_{res}=1.1\;eV$ and Gaussian
disorder $\delta=0.10\;eV$, averaging the values over a set of 500
runs. Other cases are shown in Fig. \ref{trans} (b) $42\;\%$,
$2E_{res}=1.7\;eV$, $\delta=0.19 \; eV$, (c) $24\;\%$,
$2E_{res}=2.2\;eV$, $\delta=0.26\;eV$ and (d) $7.8\;\%$,
$2E_{res}=3.7\;eV$, $\delta=0.29\;eV$. With the decrease of $sp^2$
bond concentration and increase of disorder in model A, we observe
an increase in the $2 E_{res}$ value (the effective band gap),
broadening of the peak width and a decrease in peak-to-valley ratio.
With further decrease of $sp^2$ bond concentration, the resonance
peaks move closer to the $sp^3$ band gap and finally disappear.
Models A and C show significant broadening of the resonant peaks
compared to case B for low $sp^2\;\%$ since disorder is large (see
Fig. \ref{trans}(c) and (d), insets) although it is very similar for
all cases for high $sp^2\;\%$. At large $sp^{2}$ percentages the
transmission through the structure is high therefore structural
effects only slightly influence the transmission coefficient. When
the $sp^{2}$ percentage is low, the transmission is also low hence
even slight structural changes have a large effect on $T(E)$.
\begin{figure}[t]

 \centering
      {
        \includegraphics[width=8.1675cm,  height = 5.17275cm]{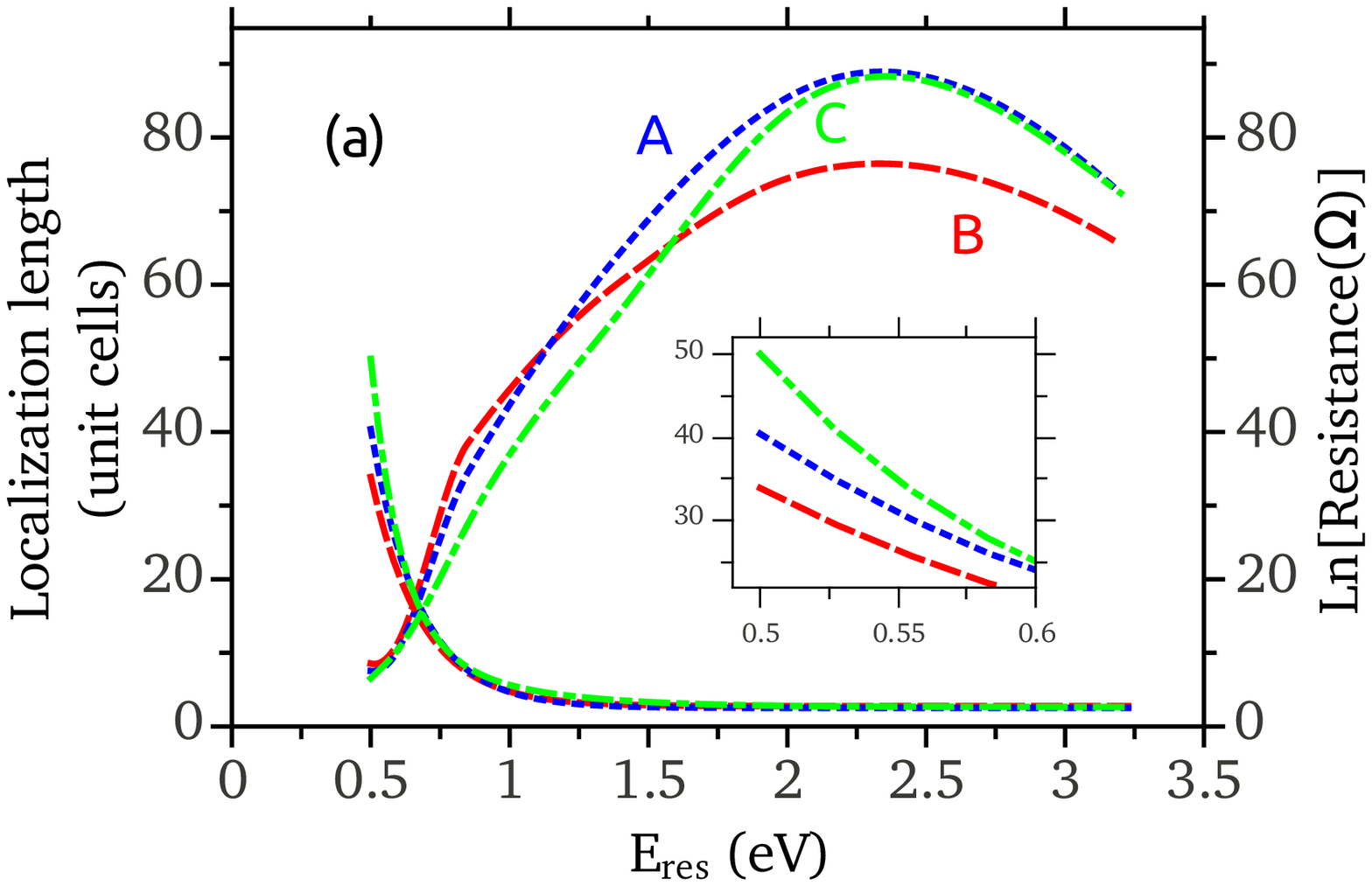}
        \label{loc_res}
    }
      {
        \includegraphics[width=6.5cm]{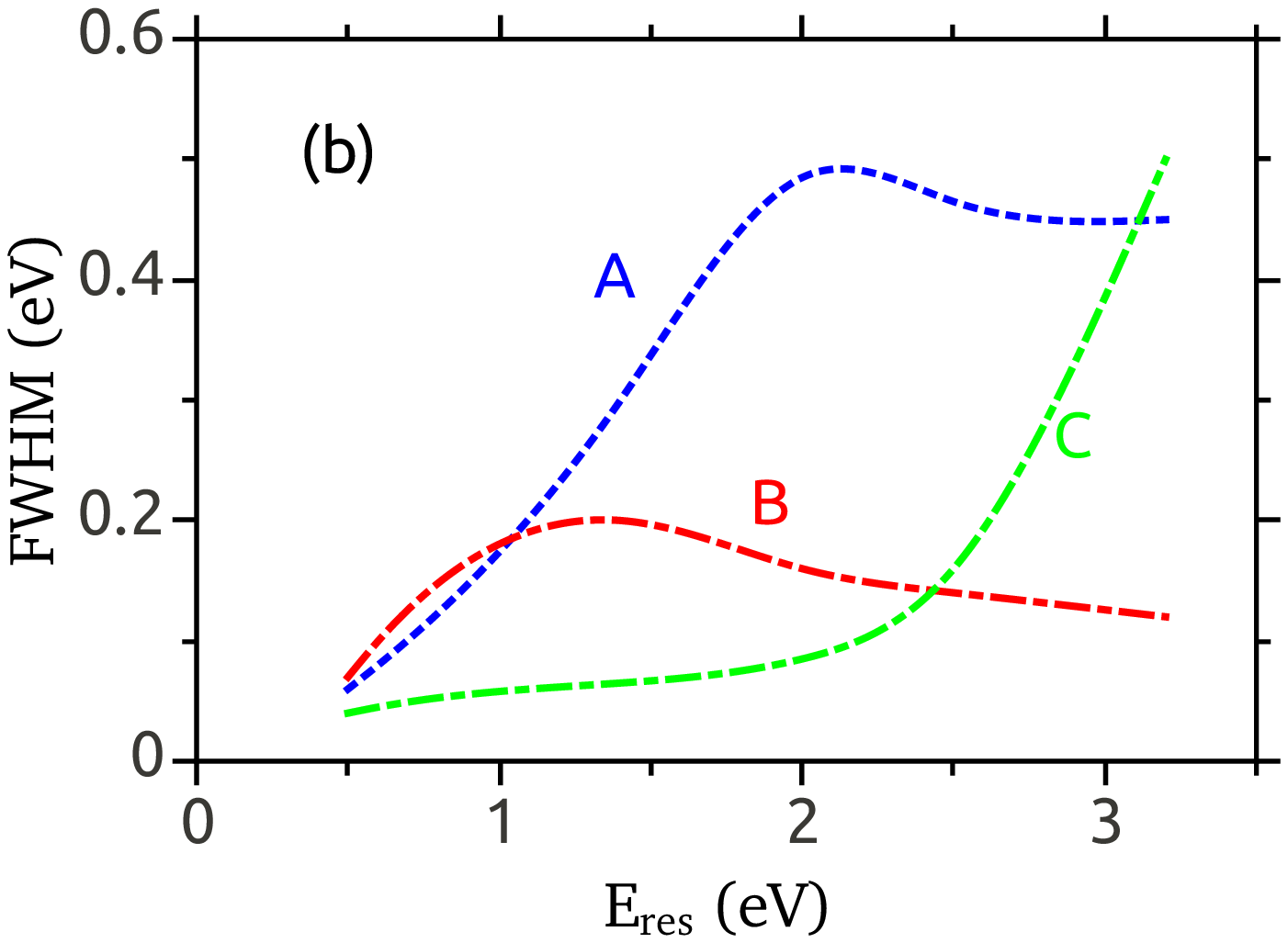}
        \label{fwhm}
    }
\caption{(a) Localization length (lower curves) and resistance (upper curves) calculated at
$E_{res}$ peaks, which are determined by the factor of disorder as
well as the energy distance from the $sp^3$ structure band edge. The
inset shows $L_{loc}$ at low energies. (b) Variation of FWHM for the
transmission peaks $E_{res}$. 3 curves (in color online) correspond
to cases A, B and C of hopping disorder parameter $\delta$
dependance on $E_{\pi}$ as shown in
Fig.\ref{pi_delta}(b).}\label{locres_fwhm}
\end{figure}

\begin{figure}[b]
\centering
\includegraphics[width=10cm, clip=true]{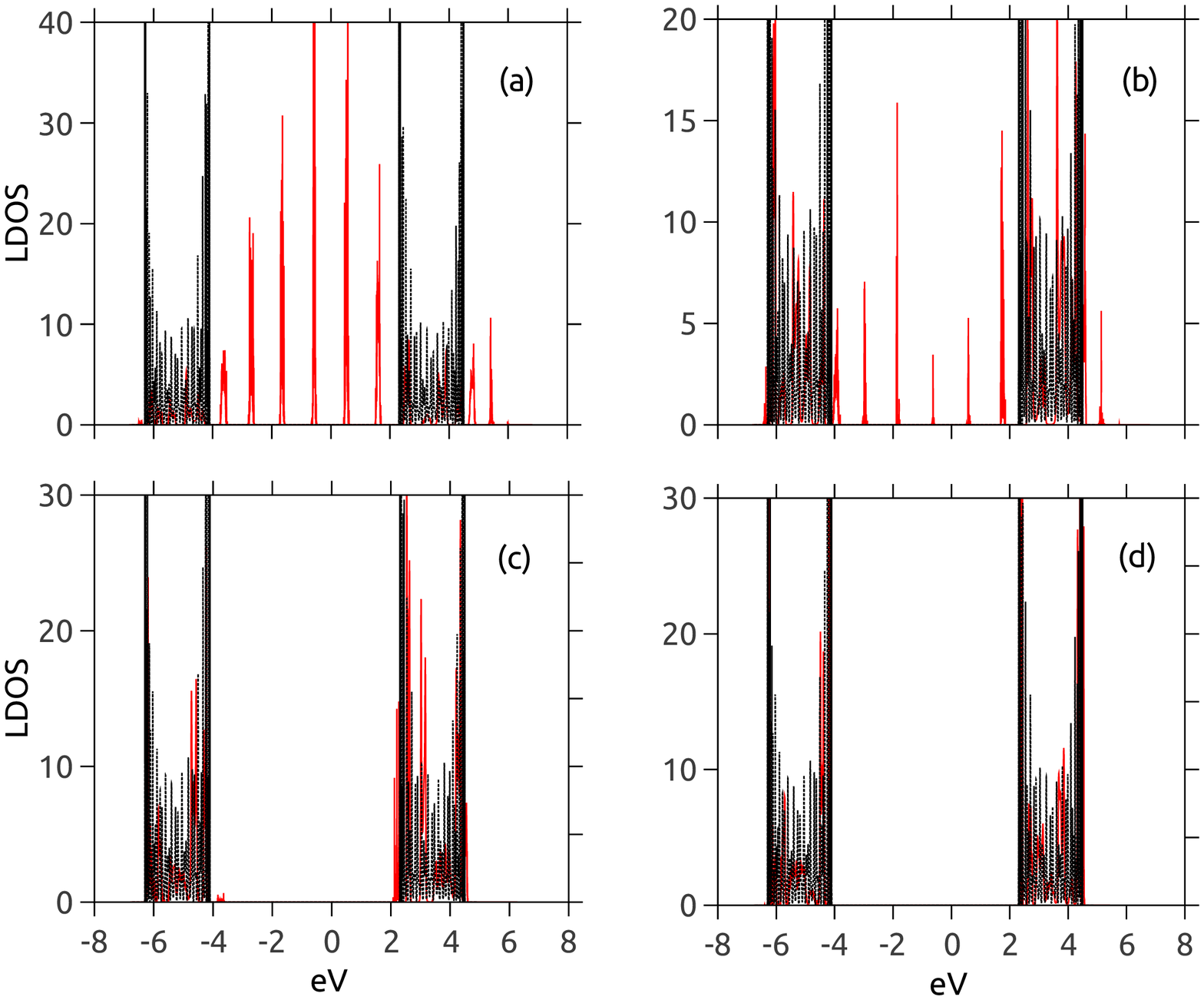}
\caption{LDOS (arbitrary units) for model A at the structure central
cites for (a) $72\%$, (b) $62\%$,(c) $42\%$ $sp^2$, and (d) $24\%$
$sp^2$ bonds (solid red curve) compared to that of pure $sp^3$
structure (dotted black curve).} \label{ldos}
\end{figure}

The disorder induces localization of the wave function within the
structure and the localization length is expressed as
$L_{loc}(E_{res})=-\frac{2L}{\ln T(E_{res})}$, where $L$ is the
length between leads used in the Landauer formula for calculating
the resistance \cite{guin2}. At small energies  $L_{loc}$ is longer
for case A and C than for case B because the structure is more
transparent at resonant energies for lower disorder
(Fig.\ref{locres_fwhm}(a), lower curves). The localization length
decreases sharply for all 3 cases till $E_{res}$ reaches the value
of about $1\;eV$ with a further saturation showing an exponential
behavior. The resistance behaves oppositely to the localization
length (Fig.\ref{locres_fwhm}(a), upper curves). At small energies
the resistance is smallest for case C and largest for case B
following the distortion parameter dependence.  The structural
resistance increases till $E_{res}= 2\;eV$ followed by a slight
decrease at higher energies in the natural logarithm scale. This
value is related to the mobility edge- as the disorder parameter
increases, the transmission maxima shift away from $E_{f}$ towards
the mobility edge. The lifetime of resonant states therefore
decreases (the full width at half maximum (FWHM) decreases). At
large energies it is greater for case A and C (almost coincide)
because of the larger distortion values. From this analysis it is
revealed that the tunnel conductance can be improved in $sp^2$-rich
carbon for model C and also in $sp^3$-rich carbon for model B.
Whereas for models A the conductance can vary as intermediately
between model B and C. Although the tunnel resistance increases
initially in the low energy region it can be controlled for a wide
band gap carbon structure. In practice the size of the $sp^{2}$
clusters can be controlled to some extent through processes such as
annealing or irradiation hence the band gap can be controlled to
some extent. Nitrogen doping (discussed below) is also a
possibility. These results explain the initial increase of
resistance of tetrahedral $a-C$ ($ta-C$) films incorporated with a
small amount of nitrogen followed by a decrease for high nitrogen
(or $sp^2-C$) concentration \cite{davis}.

 A comparative study of the FWHM for cases A, B
and C showed a significant difference not only in their absolute
values but also in the their trends as a function of energy. For
case A, the FWHM increases from low energy followed by a saturation
at energies above 2.5 $eV$. For case B, the FWHM decreases beyond
1.5 $eV$ (due to decrease of disorder). For case C, the FWHM
increases continuously above 2.5 $eV$. The other factors which
effect the FWHM are the incoming energy $E_{res}$ and the width of
the $sp^3$ ``barrier". A higher energy and smaller width increase
the peak width due to the higher transparency of the structure for
an incoming wave. The value of the FWHM, characterizing an average
quasi-bound state lifetime at a particular energy, approaches
saturation in the high $sp^3$ limit. The width of the resonant peaks
is proportional to the characteristic time of resonant states
therefore as the disorder increases, the FWHM of resonant peaks and
consequently the characteristic time of resonant states also
increases. At the same time the amplitude of transmission maxima
decreases rapidly. The conductance crossover occurs as with further
decrease of the $sp^2$ bond concentration the resonant peaks shift
towards the mobility edge and the characteristic time therefore
decreases with the resistance decreasing accordingly. The FWHM of
the resonant peaks saturate at very low $sp^2$ bond concentration
but the resistance does not saturate as the amplitude of the
resonance peaks increases as the peaks approach the mobility edge.
These results clearly establish the specific effect of the
non-linear model A, which shows the tuneable speed of carbon devices
from the slow to fast regime as the $sp^2\;\%$ increase in these
structures.

The structural change for case A is also reflected in the LDOS
accompanied by the peak position change and broadening. Fig.
(\ref{ldos}) shows the LDOS at the central cite for structures with
different $sp^2$ bond concentrations. At a high concentration (Fig.
\ref{ldos} (a)) there are  several high LDOS peaks inside
$E_\sigma^g$. As the concentration decreases, LDOS peaks decrease
being comparable  in amplitude to LDOS corresponding to the $sp^3$
bands (Fig. \ref{ldos} (b)). Finally, when the $sp^2$ bond
concentration is less than $50\;\%$ the LDOS peaks inside
$E_\sigma^g$ become negligible compared to LDOS outside (Fig.
\ref{ldos} (c) and (d)). Also, the LDOS peaks move from the zero
energy position with the conjugated phase decrease, effectively
opening the gap. The broad features of LDOS suggest the possibility
for hopping transport in these structures, which is commonly
observed in diamond-like carbon films \cite{robert,helmbold,sbprb}.
On the other hand, the high LDOS filling the $sp^3$-gap region
explains the metallic conduction in graphitic carbon films. Hence
the observed conductance crossover in $a-C$ films with the increase
of $sp^2C\;\%$ (or $N\;\%$) can be explained by the change of the
characteristic time of the resonant states \cite{sbnmat}.
\begin{figure}[t]
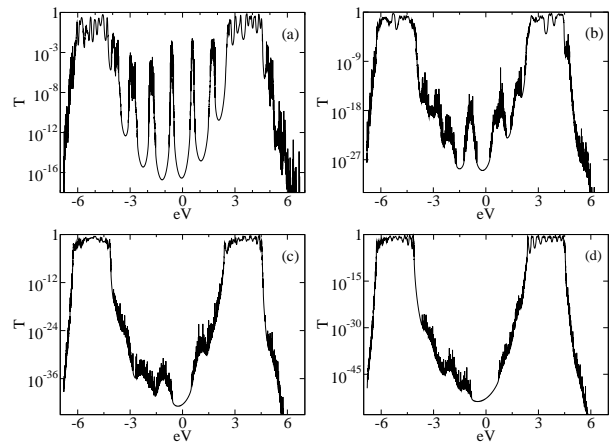

\centering
\begin{tabular}{cc}
\epsfig{file=t01.eps,width=0.45\linewidth,clip=} &
\epsfig{file=t02.eps,width=0.45\linewidth,clip=}\\
\epsfig{file=t03.eps,width=0.45\linewidth,clip=} &
\epsfig{file=t04.eps,width=0.45\linewidth,clip=}
\end{tabular}
\caption{$T(E)$ vs. $E$ for the  (a) $72\;\%$ ,(b) $42\;\%$ ,(c)
$24\;\%$, and (d) $7.8\;\%$ bond for the topological
disorder considering model A. The energy axis is drawn relative to the Fermi energy.}\label{topol}
\end{figure}

Besides the structural disorder we attempt to find the role of
topological disorder on the transport properties of the structure.
We know that topological disorder is more prominent in
two-dimensional structures e.g. large $sp^2$ clusters however, the
effects can also be simulated in quasi-one dimensional (small $sp^2$
clusters) structures. In the topologically disordered case for
carbon superlattices or heterostructures, non-uniform $sp^2-C$
cluster sizes were distributed throughout the system in contrast to
the non-topologically disordered case where all $sp^2$ clusters have
the same size. Thus, it corresponds to so-called Anderson vertical
disorder. The disorder parameter was determined using model A. As it
can be seen from the Fig. \ref{topol}, the topological disorder
results in the decrease of the peak height and increase of the
peak width and has a greater effect on the peaks located further
from the Fermi level. The effect is more significant for structures
with lower $sp^2$ concentration, so the relative change in the
quasibound energy levels is high. In addition to that, new minor
peaks appear in when the difference in energy levels induced by
topological disorder is larger than the hopping structural disorder
parameter. Topological disorder modifies the resonant tunneling
conditions, changing the resonant energy conditions of consecutive
wells resulting in a mismatch of the resonant levels. The amplitude
of transmission peaks at resonant energies therefore decreases and
the resonant peaks spread out with energy as resonance no longer
occurs at a sharply defined energy. A resonance peak splits in subpeaks
corresponding to the cluster number (4 in this work), where
topological characteristic energy splitting $\delta
\varepsilon_{top}$ is proportional to $\frac{\delta w}{w^3}$, where
$w$ and $\delta w$ are  the average cluster size and its
characteristic variation. When $\delta \varepsilon_{top}$ is larger
than the characteristic structural disorder variation determined via
deformation potential as $\delta\varepsilon_{str}=\gamma\delta l$,
where $\gamma$ is the deformation potential and $\delta l$ is the
bond length variation, the subpeaks are distinguishable. With
increasing structural disorder the subpeaks start to spread out and
finally overlap when $\delta\varepsilon_{top}$ becomes smaller than
$\delta \varepsilon_{str}$ with the elimination of the distinct
subpeaks features due to the fact that the number of clusters is
much smaller than the number of sites. The resonant peaks can therefore be
tuned by disorder of both types which can be controlled to some extent
experimentally through annealing or irradiation of the films
\cite{sb-nexafs}. Based on the bond length distribution included in
the model, different ring structures e.g. five fold and seven fold
symmetry, additional so-called topological disorder can be included
subject to further study.

\textit{Conclusion}: In this Letter, the importance of the
non-linear disorder vs. energy model (case A), an intermediate to
uniform increase and decrease of disorder, is established. The
theoretical analysis of this experimentally supported model
discovers the possibility for a relative increase of the tunnel
conductance in $sp^3C$-rich carbon due to the appearance of resonant
transmission peaks close to the bands. Nitrogen doping, for example,
could be used to vary the micro-structure of $a-C:N$ films as
nitrogen incorporation increases the $sp^2-C$ ratio \cite{davis,
sb-nexafs}. The nitrogen concentration also influences the disorder
\cite{chimowa}. Nitrogen doping is therefore not the same as
conventional doping however it can still be exploited to modify the
conductivity of $a-C$ films \cite{davis}. In this regard the result
of the present study can be extended for doped carbon structures. A
detailed study of nitrogen incorporation in $a-C$ structures has
recently been carried out based on a model related to this work
\cite{katkov}. The lifetime of electrons can be nearly constant over
a large range of $sp^3C\;\%$ in carbon structures having a wide
bandgap energy. Since the intensity and position of the resonant
peaks can be tuned by the $sp^2C\;\%$ to $sp^3C\;\%$ ratio and
associated disorder under this model, we can effectively predict the
nano-electronic device properties in a wide range of undoped and
doped carbon structures. This model can also explain the origin of
observed resonant features in multi-layered carbon systems
\cite{sbnmat}.

The work was supported by the URC(WITs) and the NRF(SA) under the
Nanotechnology Flagship Project. We would like to thank R. McIntosh for discussion.


\begin{thebibliography}{99}
\bibitem{beeman} D. Beeman, J. Silverman, R. Lynds, and M. R. Anderson, Phys. Rev. B {\bf 30}, 870 (1984).
\bibitem{robert} J. Robertson and E. P. O'Reilly, Phys. Rev. B {\bf 35}, 2946 (1986).
\bibitem{bred} J. L. Bredas and G. B. Street, J. Phys. C, {\bf 18}, 651 (1985).
\bibitem{tersoff} J. Tersoff, Phys. Rev. Lett. {\bf 61}, 2879 (1988).
\bibitem{galli} G. Galli, R. M. Martin, R. Car, and M. Parrinello., Science {\bf 250}, 1547 (1990).
\bibitem{godet} C. Godet, Diamond Relat. Mater. {\bf 12}, 159 (2003).
\bibitem{kugler} S. Kugler and I. Laszlo, Phys. Rev. B {\bf 39}, 3882(1989).
\bibitem{dasgupta} D. Dasgupta, F. Demichelis, C. F. Pirri, and A. Tagliaferro, Phys. Rev. B {\bf 43}, 2131 (1991).
\bibitem{mckenzie} D. R. McKenzie, D. Muller, and B. A. Pailthorpe, Phys. Rev. Lett. {\bf 67}, 773 (1991).
\bibitem{kelires} P. C. Kelires, Phys. Rev. Lett. {\bf 68}, 1854 (1992).
\bibitem{frauen}Th. Frauenheim, P. Blaudeck, U. Stephan, and G. Jungnickel, Phys. Rev. B {\bf 48}, 4823 (1993).
\bibitem{tamor} M. Tamor and W. Vassel, J. Appl. Phys. {\bf 76}, 3823 (1994).
\bibitem{stumm} P. Stumm, D. A. Drabold, and P. A. Fedders, J. Appl. Phys. {\bf 81}, 1289 (1997).
\bibitem{ferrari} A. C. Ferrari and J. Robertson, Phys. Rev. B {\bf 61}, 14095 (1999).
\bibitem{franch} G. Fanchini and A. Tagliaferro, Appl. Phys. Lett. {\bf 85}, 730 (2004).
\bibitem{cherkashinin} G. Cherkashnin, O. Ambacher, T. Sciffer, and G. Schmidt, Appl. Phys. Lett. B {\bf 88}, 172114 (2006).
\bibitem{carey} J. D. Carey and S. R. P. Silva, Phys. Rev. B {\bf 70}, 235417 (2004).
\bibitem{davis} J. Robertson, and C. A. Davis, Diamond. Relat. Mater. {\bf 4}, 441 (1995).
\bibitem{alibart} F. Alibart, M. Lejeune, O. Durand Drouhin, K. Zellama, and M. Benlahsen, J. Appl. Phys. {\bf 108}, 053504 (2010).
\bibitem{shimakawa} K. Shimakawa and K. Miyake, Phys. Rev. Lett. {\bf 61}, 994 (1988).
\bibitem{helmbold} A. Helmbold, P. Hammer, J. U. Thiele, K. Rohwer and D. Meissner, Philos. Mag., B,  {\bf 72}, 335 (1995).
\bibitem{sbprb} S. Bhattacharyya, Phys. Rev. B {\bf 77}, 233407 (2008).
\bibitem{mikulski} P. Mikulski, J. Patyk, and F. Rozploch, J. Non-Cryst. Solids {\bf 176}, 230 (1994).
\bibitem{guin2} I. Martin and Ya. M. Blanter, Phys. Rev. B {\bf 79}, 235132 (2009).
\bibitem{sbnmat} S. Bhattacharyya,  S. J. Henley,  E. Mendoza,  L. Gomez-Rojas,  J. Allam and S. R. P. Silva, Nature Materials {\bf 5}, 19 (2006).
\bibitem{onipko} A. Onipko, Phys. Rev. B {\bf 59}, 9995 (1998).
\bibitem{Efstathiadis} H. Efstathiadis, Z. Akkerman, and F. W. Smith, J. Appl. Phys. {\bf 79}, 2954 (1996).
\bibitem{Stephan} U. Stephan, Th. Frauenheim, P. Blaudeck, and G. Jungnickel, Phys. Rev. B {\bf 50}, 1489
(1994).
\bibitem{hjort} M. Hjort and S. Stafstr\"{o}m, Europhys. Lett. {\bf 46}, 382 (1999).
\bibitem{sb-nexafs} S. Bhattacharyya, M. Lubbe, P. R. Bressler, D. R. T. Zahn, and F. Richter, Diamond. Relat. Mater. {\bf 11}, 8 (2000).
\bibitem{chimowa} G. Chimowa, D. Churochkin, and S. Bhattacharyya, Europhys. Lett. in press (2012).
\bibitem{katkov} M. V. Katkov and S. Bhattacharyya, J. Appl. Phys. {\bf 111}, 123711 (2012).




\end{thebibliography}
\end{document}